\documentclass[runningheads]{svmult}

\usepackage{makeidx}   
\usepackage{graphicx}  
\usepackage{subeqnar}  
\usepackage{multicol}  
\usepackage{physprbb}  
\makeindex             

\def\be{\begin{equation}}
\def\ee{\end{equation}}
\def\bea{\begin{eqnarray}}
\def\eea{\end{eqnarray}}
\def\cf{{\it cf.}}
\def\eg{{\it e.g.}}

\def\etal{{\it et al.}}

\def\go{\mathrel{\raise.3ex\hbox{$>$}\mkern-14mu\lower0.6ex\hbox{$\sim$}}}
\def\lo{\mathrel{\raise.3ex\hbox{$<$}\mkern-14mu\lower0.6ex\hbox{$\sim$}}}
\begin{document}
\title*{To the Lighthouse}
\toctitle{To the Lighthouse}
\titlerunning{To the Lighthouse}
\author{R. D. Blandford}
\authorrunning{R. D. Blandford}
\institute{Caltech, Pasadena, CA 91125}
\maketitle
\begin{abstract}
The extreme hypothesis that essentially all types of ultrarelativistic 
outflow -- specifically AGN jets, pulsar wind nebulae 
and GRB --are electromagnetic,
rather than gas dynamical, phenomena is considered. 
Electromagnetic flows are naturally 
anisotropic and self-collimating so as to produce 
jet-like features. The
relativistic force-free description of these flows, which is simpler than the
relativistic MHD description, is explained. It is suggested that the 
magnetic field associated with AGN jets and GRB is quite extensively
distributed in latitude, 
without necessarily increasing by much the total power. 
It is also proposed that the observed 
emission from these sources
traces out regions of high current density where global 
instabilities drive a turbulence spectrum that is ultimately 
responsible for the particle acceleration and the synchrotron,
inverse Compton and synchro-Compton
emission. The direct
extraction of spin energy from a black hole is re-examined and an
electromagnetic model of GRB explosions is 
developed. It is also suggested that some GRB ``lighthouses''
be identified with accretion-induced collapse of a neutron star to form a 
black hole in a binary system.
\end{abstract}
\section{Introduction}
For an international audience, I should explain that my title is 
taken from a novel written by Virginia Woolf. It is one of the 
premier examples of the ``stream of consciousness'' school of narrative 
fiction, a style that lends itself well to scientific conference proceedings.
If I take literally the metaphor that 
inspired this meeting, lighthouses draw attention to themselves 
(and warn of incipient danger) by 
shining brightly through obscuring material and varying,
just like their cosmic counterparts. Now, nearly
all of the variable, cosmic beacons 
that we have discussed here have, implicitly,
involved the formation or activation of a black hole. However, few 
of the other talks have remarked upon this fact, let alone analyzed
its physical implications. To most observers and phenomenologists, 
a black hole might just as easily be a black box - a flexible
source of power whose properties are
limited only by its total mass and the referees
of theoretical papers. It is my task to ``deconstruct'' the lighthouse. 

I must first emphasize an important point.
Provided that general relativity is the correct theory of strong gravity,
(and there is a great need to test this proposition directly),
we have a very good understanding of the theoretical 
properties of black holes. The Kerr spacetime, the stage upon which
epic cosmic dramas are enacted, introduces distinctive but quite calculable 
modifications to fundamental physics which we may even be observing
directly. Our interpretation of these observations remains quite sketchy,
on account of our poor understanding of the nineteenth century subjects of 
gas dynamics, electromagnetism and statistical mechanics, not the general
relativity and quantum mechanics of the twentieth!

In this article, I will focus on two linked issues. The first involves 
revisiting the extraction of black hole spin energy 
through the application of electromagnetic torques.
Here general relativity is of crucial importance.
The second is to consider the time-dependent, special relativistic
evolution of a magnetic shell containing toroidal
magnetic field that has been wound up by a central, spinning, compact object,
specifically a black hole plus accretion disk or a neutron star.
I will report on an analysis being carried out with Max Lyutikov
and argue that this model provides an alternative explanation for 
gamma ray bursts and is clearly contrasted with the standard, fireball
picture. These considerations are also relevant to pulsar wind nebulae (PWN) 
and possibly to relativistic jets in AGN. 
\section{Ultrarelativistic Outflows}
\subsection{AGN}
\subsubsection{Jets}

We have known, since the earliest days of VLBI, that quasars create 
ultrarelativistic outflows\cite{ree67} \cite{coh71} with Lorentz factors 
$\Gamma\sim10$ and that this flow
is typically collimated into two antiparallel jets with opening
angles $\sim5^\circ$. We know now that 
the radio emission is only the smoke
\cite{bri84}. The $\gamma$-ray fire \cite{har92} is far 
more powerful and originates from smaller radii than the radio
waves. The $\gamma$-rays are believed to be produced by inverse
Compton scattering (of scattered, disk photons in the case of high 
power sources and jet synchrotron photons in the case of low power sources
\cite{pad00}).
$\gamma$-rays with energies as 
high as $\sim1$~TeV have been recorded, with variability as rapid
as 15 min \cite{gai96}. 

Although jets are commonly modeled as stationary outflows, they are 
quite variable and this variation can be 
observed at radio wavelengths using VLBI. 
The phenomenon of superluminal expansion 
is usually attributed to outwardly propagating, 
relativistic, internal shocks, that are caused by outbursts
close to the central black hole \cite{bla79}. 
The kinematics of these shocks can be quite complex
and many intricate models of their emission properties have been constructed.
The energies associated
with individual outbursts, lasting of order a week,
can be as high as $\sim10^{54}f$~erg, where $f\sim10^{-3}$ 
is the beaming fraction.
\subsubsection{Disk and Holes}   

The lighthouses are the central
black holes with masses up to $\sim10^9$~M$_\odot$
and their attendant accretion disks \cite{geb00}, \cite{fer01}. 
In most scenarios, the radiant power derives from the binding 
energy of the accreted gas. Formally this is limited by the existence of a 
smallest, stable, circular, Keplerian orbit and ranges from 
$\sim0.5-4\times10^{20}$~erg g$^{-1}$ as the hole angular velocity changes 
from 0 to $0.5/m$. This energy channel 
is unavoidable while black holes are
feeding voraciously and growing rapidly and it is probably responsible for
most of the AGN luminosity density in the universe (although the details
of how the photons are produced remain unclear
\cite{ago01}). However, it is far from obvious that it can account for
ultrarelativistic outflows. The main objection is that gas dynamical
winds from gravitationally bound reservoirs, like the solar wind
and bipolar outflows from young stellar objects, usually have terminal 
speeds that are not much more than the gravitational escape velocity
at their footpoints
\cite{kon99}, whereas AGN jets are clearly ultrarelativistic. 

There are several ways out. One hypothesis is that, instead of being radiated
efficiently, the binding energy is dissipated in a corona as a high entropy
-- radiation and pair-dominated fluid that forms the base of
a thermally-driven, hot wind. As this fluid expands, the
radiation decompresses and the pairs annihilate, just like in the expanding
universe, leaving the flow baryon-dominated.  
The difficulty with this hypothesis, in the 
case of AGN, is that most of the prominent radio jets that we observe
seem to be relatively low entropy sources. For example, 
in the radio galaxy M87 (which forms jets within $\sim100m$ \cite{jun99}), 
the hole mass is $M\sim3\times10^9$~M$_\odot$ while the bolometric
luminosity appears to be $L\sim10^{43}$~erg s$^{-1}\sim3\times10^{-5}
L_{{\rm Edd}}$, and is probably less than the jet power. (This objection
does not apply to the ``Galactic Superluminal Sources'', which do appear
to be formed in radiation-dominated environments and only have 
mildly relativistic speeds.) 

Another hypothesis is that the working substance is a pure,
electron-positron pair plasma. However, this is problematic
because radiative drag in the nucleus
precludes acceleration to more than mildly relativistic speed.
Yet another possibility is that the jet momentum is carried 
by ultrarelativistic protons, which may have been accelerated by shock fronts 
formed close to the black hole. There are two difficulties 
here. The first is that VLBI polarization observations are best interpreted
in terms of a pair plasma \cite{war98}. The second is in collimating the
outflow. This is effectively impossible using gas pressure without
violating observational constraints on the X-ray luminosity. (Electro)magnetic
field with an energy density comparable with the protons therefore have to be 
invoked to effect the collimation under the conditions thought to 
obtain in AGN. 

For these and other reasons,
many astrophysicists have suggested that AGN jets are powered
and collimated hydromagnetically, as also appears to 
be the case in protostellar jets. A variety of detailed mechanisms have been
proposed \cite{bla00}. 
In most of these mechanisms, the magnetic field derives 
from the gas in the accretion
disk. The field lines may be primarily poloidal, they may be toroidal 
or they may involve poloidal field becoming toroidal with increasing 
distance from the disk. The magnetic field may have large scale order over
many octaves of radius \cite{kon99} or be quite tangled
\cite{tou96}. When the disk orbits 
with a near-Keplerian velocity, it is possible to launch gas centrifugally 
\cite{kra99}
and, although it may be possible to create an ultrarelativistic terminal 
velocity, the natural presumption is that the terminal velocity
of an outflow emanating from an accretion disk
is no more than mildly relativistic. 

These considerations have, in turn, motivated the investigation of an 
alternative possibility, that the 
power and speed of the jets follow from their direct magnetic connection
to the spinning black hole. (The collimation can still be caused by 
a mildly relativistic, disk outflow, though.)  Black holes possess rotational
energy -- the difference between the mass $m=m_0(1-\beta^2)^{-1/2}$ and the 
irreducible mass $m_0$,  where $\beta=2m_0\Omega_H<2^{-1/2}$, with
$\Omega_H$ being the hole angular velocity
\cite{tho86}. Up to 29 percent of the mass 
is, in principle, available.
For a billion solar mass hole this is $\sim6\times10^{62}$~erg,
ample for the most prolifigate of extragalactic radio sources.
Furthermore, connecting the jet to the event horizon makes 
it very difficult for  
plasma from the disk to attach itself to these field lines and it 
is natural to produce an ultrarelativistic, baryon-starved outflow
under these conditions. We explore this idea further below. 
\subsection{Pulsar Wind Nebulae}
\subsubsection{Pulsars}

Pulsars were first discovered through their 
pulsed radio emission and quickly identified as spinning, magnetized neutron
stars. More recently, some of them have been identified as far more powerful
$\gamma$-ray sources. However most of the energy that they release is 
in the form of a relativistic outflow which 
inflates a pulsar wind nebula, like the famous Crab Nebula. The outflow 
Lorentz 
factors in pulsar winds have sometimes been estimated to exceed $\sim10^6$
and it has been argued that they become strongly particle-dominated by the 
time they reach the nebula \cite{ken84}. 
On this view, which is like a high speed
version of the fluid model
of AGN jets, there is a transformation of the power from Poynting flux
to bulk kinetic energy of a fluid at fairly small radius,
so that a strong, fluid shock can be produced 
when the momentum flux in the outflow balances the ambient pressure
in the nebula. However, recent X-ray observations
of the Crab Nebula exhibit strong synchrotron emission concentrated along 
the axes and the equatorial plane \cite{wei00} and there 
is little evidence for a strong shock at intermediate latitude, as expressly
proposed in the fluid model. These 
observations encourage us to re-examine
magnetic models of PWN. 
\subsubsection{Electromagnetic Model of the Crab Nebula}

The Crab pulsar is a spinning, magnetized, 
neutron star with an inclined dipole 
moment. It presumably has a force-free magnetosphere through which
currents flow. The complete electrodynamical
description of this magnetosphere remains an unsolved problem. However,
it seems plausible that,
somewhere beyond the light cylinder, the electromagnetic
field becomes essentially axisymmetric and that variation on the scale of a
wavelength dies away. There are at least three ways by which this can occur.
There can be steady reconnection in the outflowing, ``striped'' wind
\cite{uso94}, \cite{lyu01}.
Alternatively, the waves can decay successively 
through parametric instability
into higher frequency waves
\cite{max72}. These two processes are essentially 
dissipative and the magnetic energy will ultimately be converted into 
heat. However, a third option is essentially 
non-dissipative. Consider the minority of the open magnetic field lines
that emanate from the neutron star's southern
magnetic pole, and which can be followed into the northern hemisphere.
These may be pulled back across the equatorial plane 
by magnetic tension into the southern hemisphere
(and {\em vice versa}).  Any one of these
three mechanisms will destroy the AC component of the electromagnetic
field leaving behind only the DC component in which the magnetic
field will become increasingly toroidal and axisymmetric with radius.
If the magnetic field develops this structure, there must be an associated
current flow out along the poles and eventually returning in the equatorial
plane, or {\em vice versa}. 

Now consider what must happen to this relativistic,
magnetic wind. As the PWN expands with 
a speed of only $\dot R\sim1000{\rm km s}^{-1}<<c$, where $R(t)$
is the radius of the bubble, it is easy to see that the pulsar must be 
producing magnetic flux at a rate that is roughly $(c/\dot R)^{1/2}$ 
times too large to account for the strength of the magnetic 
field in the nebula. Therefore, most (95 percent in the case of the 
Crab Nebula) of the flux 
must be destroyed
\cite{ree73}. On topological grounds, the natural
places for this destruction to occur are on the axis and the equatorial
plane \cite{beg98} and these regions are, in any case, 
formally unstable to pinch and tearing mode instabilities.
(The contact discontinuity, although formally stable to 
Kruskal-Schwarzschild modes is, in practice likely to be unstable to other 
hydromagnetic instabilities.) There must be
a fairly rapid migration of magnetic flux towards the poles and the 
equator, where it will become tangled on progressively
smaller lengthscales in a turbulence spectrum
until it can reconnect at a resistive inner scale. Looking at the
same process from a current point of view, in the context of the Crab Nebula,
we can imagine that the X-ray synchrotron emission delineates regions where a
quadrupolar current flow becomes dissipative as a consequence
of these instabilities and accelerates relativistic
electrons and positrons which diffuse outwards into the body of the nebula.
(The current flow observed within the heliosphere, as infered from observations
by the Ulysses spacecraft, is also mostly confined to the axis 
and the equator.) 

What would happen if we were to prevent the dissipation required in the 
Crab Nebula through having the current flow along rigid, 
perfect conductors? The answer turns out to be that the Poynting
flux would be reflected by the outer boundary of the PWN and channeled
back onto its pulsar source where it would react back upon the power supply.
\subsubsection{Magnetars}

Magnetars are most convincingly interpreted as slowly rotating
($P\sim5-10$~s, high field $B\go10^{14}$~G pulsars)
\cite{kou99}. Most of the energy
that they lose is probably derived from the internal magnetic field and
released explosively \cite{spr99}, \cite{tho02}. 
The timescales for energy release are almost surely 
much longer than the light crossing times of the magnetosphere and they will
form a complex, anisotropic, electromagnetic pulse that expands essentially
at the speed of light.  As the pulsar spin period is small, the 
pulse carries essentially no angular momentum, just like a model of 
a pulsar wind in which we ignore the poloidal magnetic field.  
\subsubsection{Implications for AGN Jets}

In the above model of the Crab Nebula, the dynamics is dominated by 
electromagnetic field everywhere within the contact
discontinuity that separates the PWN from the shocked
interstellar medium; the inertia and the pressure of the 
plasma is of minor importance.
(We actually know that some thermal plasma does evaporate off the
expanding debris from the original supernova explosion so it is not 
totally ignorable in practice
\cite{wil74}.) Can we imagine that the same is true of AGN jets?
Can it be that the ``jet'' features that we image at radio, optical
and X-ray wavelengths are really just delineating the regions 
of high current density? In other words, are these jets surrounded
by extensive, evacuated sheaths of toroidal field that flow
radially inward to compensate the flux that is destroyed as a result
of electromagnetic instabilities
\cite{bla76}, \cite{lov76}, \cite{ben78}? 

One of the merits of this
viewpoint is that it provides a dynamical rationalisation of the 
doctrine of equipartition. Pinched currents, on all scales, may continue
to become unstable until their stresses are balanced by pressure. 
Another merit is that it provides a 
natural explanation for the helical structures that are often seen
in VLBI maps. A third advantage is that currents can account
for distributed  particle acceleration in well-resolved jets, as 
recent spectral 
studies suggest may be required \cite{wil02}.
\subsection{GRB}
The most interesting and the most pressing challenge, though, is to 
understand $\gamma$-ray bursts (GRB)
\cite{mes01}. Most contemporary models invoke
a high entropy fireball which eventually creates
a baryon-dominated, fluid outflow as discussed above. Internal shocks
in this outflow, derived from source variability, are responsible for the 
$\gamma$-ray burst itself; the relativistically expanding shell of 
shocked interstellar medium being the site of the afterglow. The source
itself is most popularly associated with ``hypernovae'' or ``collapsars''
\cite{mac01} 
-- evolved, rapidly rotating massive stars that form  
a ``twin exhaust'' \cite{bla74} 
through which a pair of ultrarelativistic, fluid jets escape. The best
evidence in support of this model is the observation of achromatic breaks
in some afterglows. These are expected to arise when the expansion Lorentz
factor falls to a value equal to the reciprocal of the jet opening
angle. Put another way, the emission declines more rapidly after a sound wave 
can cross the jet on the expansion timescale.   

However, it is possible to consider electromagnetic models 
\cite{uso94} in the case
of GRB as well - cold, steady state models rather than hot big bangs!
In other words, the electromagnetic view is that GRB are much more like the
other ultrarelativistic flows that we observe than the early universe.
(It is possible that the reason that the fireball
interpretation took hold in the context of GRB and not in 
AGN jets, is that the former were first discovered and studied as
$\gamma$-ray sources whereas the latter were initially observed as
radio sources.)

There are several reasons for considering electromagnetic models. 
Perhaps the strongest argument is that
the outflow has to have a very large Lorentz factor, $\Gamma\sim300$
\cite{lit01}, 
in order to reduce the pair production opacity for the 
highest energy $\gamma$-rays to a value below unity at a radius where
internal shocks can still operate. This implies that the ratio 
of bulk kinetic energy to internal energy in the jet is 
$\go10^5$. This is unprecedented in gas dynamics.
It is extremely hard
to create hypersonic, high Mach number flow in the laboratory using a carefully
machined nozzle. There are always transient rarefactions and compression
waves associated with the walls that cause large fluctuations in 
the velocity field and subsequent heating.

It should be noted that the relatively well-studied long 
duration bursts last for $\sim10^6$ source light crossing 
times, if they really are associated with the formation of a
$\sim10$~M$_\odot$ black hole. (By contrast we have 
only observed quasars for $\sim10^5$ light crossing times!)
The sources are quasi-steady on the small scale just like AGN.
It is therefore possible that the transport of energy from the source
to the $\gamma$-ray emission region and the expanding blast wave
is entirely electromagnetic in this case as well.  
There are further advantages of the electromagnetic model, 
which we discuss further below. As the dissipation is 
initiated by electromagnetic instabilities, the $\gamma$-ray emission region
does not have to be tied to the central source through internal
shocks and can, consequently,
be located much further away, where the Lorentz factor, beaming
and pair production opacity 
can be less extreme, as long as the 
variations do not average out to produce a smooth burst
Another advantage is that the electromagnetic field
is naturally self-collimating and creates an anisotropic explosion. It is 
not necessary to invoke collimating channels as in fluid models. A third
benefit is that the internal sound speed is effectively
$c$ rather than $3^{-1/2}c$ as in the fluid model. 
This implies that there is more causal contact
across the blast wave when it is pushed electromagnetically. However,
it also implies that shocks are not likely to be responsible for 
the emission of $\gamma$-rays.
\subsection{Electromagnetic Lighthouses}
We have already identified massive black holes as the lighthouses for 
AGN jets and neutron stars as being responsible for PWN. For GRB,
we cannot see the sources directly (though this could change with the advent
of neutrino or gravitational radiation observations). Consequently 
there are many possibilities. In the context of the hypernova model,
it can be envisaged that the fireball is created by large scale 
electromagnetic interactions involving the newly-formed, spinning black hole
and its surrounding torus. It is commonly supposed that the energy 
is quickly transformed into a fireball and a baryonic jet
\cite{mes01}.
The major concern with this hypernova explanation is whether the baryons
can be excluded efficiently from the flow, especially where it 
becomes trans-sonic, so that the jet can achieve as large a 
Lorentz factor as is inferred.

Now, the model that I am discussing here is initially the same as at least
some versions of the hypernova model that I have just described
\cite{lee00}. However it departs from it by assuming that the energy 
remains in an electromagnetic form all the way out to the region where
the $\gamma$-rays are emitted. There is no fireball or hadronic intermediate
state. It also deviates from other 
electromagnetic models by assuming that it is only the DC field component 
that is important well beyond the light surface.
Now it is possible that all this could happen inside a star and 
that the entrainment of gas from the star into the electromagnetic 
outflow could be negligible. However, as this does seem rather improbable,
at least to this reviewer, it is worth considering some alternative
choices of prime mover for GRB. 

One possibility is that a GRB is a newly formed magnetar
\cite{dun92} -- a strongly 
magnetised, rapidly spinning, neutron star that is able to blow away 
the surrounding stellar envelope before it slows down. To be specific,
for a long duration GRB, if the star is to have 
an electromagnetic power of $\sim10^{49}$~erg s$^{-1}$
for a time $t_{{\rm source}}\sim100$~s, then a period of $\sim4$~ms
and a field $\go10^{15}$~G is necessary at breakout. Alternatively, a GRB
may be formed as a result of the operation of an r-mode instability in a
spun-up neutron star \cite{spr99}.

Another possibility is accretion-induced neutron star collapse
\cite{mac99}.
Suppose that, a neutron star in a binary system accretes gas from its 
companion over the course of time and is able to accept the additional 
$\sim0.4-1$~M$_\odot$ required to exceed the Oppenheimer-Volkov
limit, in much the same way that white dwarfs are now thought to grow
to become SNIa. Suppose that not all of the mass in the star 
crosses the event horizon and $\sim0.1$~M$_\odot$ of gas is left behind in 
a relativistic torus, as a result of neutrino or centrifugal
stresses, and that this torus traps a flux $\sim10^{26}$~G cm$^2$. 
The spin energy of the black hole may be extracted to release the 
requisite electromagnetic energy over the observed timescale.
One advantage of this model is that there will be 
no stellar envelope to impede the escape of an ultrarelativistic outflow.

A further idea, that is only credible for the long duration GRB, is that
the source be a massive ($\sim10^5$~M$_\odot$) black hole that 
captures a white dwarf which is tidally disrupted and generates
a large magnetic field $\sim10^{10}$~G when its orbit becomes
relativistic
\cite{bla02}. In this case, the electromagnetic energy must be released 
on an orbital timescale. (It is in principle possible to release
an energy $>>M_\odot c^2$ by tapping the spin energy of the hole in this 
type of model.)
 
For the remainder of this article I shall explore the 
extreme hypothesis that 
all ultrarelativistic outflows are essentially electromagnetic
rather than fluid in character and that the dissipating/accelerating
regions coincide with the most intense current densities rather than 
shock fronts
\cite{lev97}. I shall emphasize the complementarity between 
field and the current formulations.
As most of this will be described in greater detail
elsewhere, I shall mostly confine my attention to two key aspects --
the direct extraction of energy from a spinning hole and the dynamics
of GRB. 
\section{Some Formal Preliminaries}   
\subsection{Force-free Formalism}\label{subsec:formal}
The simplest way to describe electromagnetic field both in the 
vicinity of a black hole event horizon and in an electromagnetic 
outflow is using the force-free approximation. In a flat space,
this amounts to solving the two Maxwell
equations:
\bea
\label{twomax}
{\partial\vec E\over\partial t}&=&\nabla\times\vec B-\mu_0\vec j\\
{\partial\vec B\over\partial t}&=&-\nabla\times\vec E
\eea
where the current density perpendicular to the local magnetic field
is determined by the force-free condition, which drops the inertial 
terms
\be
\label{forcefree}
\rho\vec E+\vec j\times\vec B=0
\ee
(Note that this implies that the invariant 
$\vec E\cdot\vec B=0$ and that electromagnetic energy
is conserved, $\vec E\cdot\vec j=0$.)

Under stationary and axisymmetric conditions, these equations guarantee
that the angular velocity $\Omega$ is conserved along field lines.
They also require a space charge density $\rho=\mu_0^{-1}\nabla\cdot\vec E$
of magnitude $\sim\Omega B$ to develop. (Formally this, like the equation
$\nabla\cdot\vec B=0$, is just an initial condition.)

The force-free condition can be re-expressed by setting the divergence of the 
electromagnetic stress tensor to zero
\cite{lan75}. This form has the merit that 
it brings out the analogy with fluid mechanics. Electromagnetic stress 
pushes and pulls electromagnetic energy which moves with 
an electromagnetic velocity $\vec E\times\vec B/B^2$, perpendicular
to the electric and magnetic fields. This is the velocity of the
frames (only defined up to an arbitrary Lorentz boost along
the magnetic field direction)
in which the electric field vanishes, (provided that the other invariant
$B^2-E^2>0$).
 
If we combine Eq.~(\ref{forcefree}) with the equation
\be
{\partial(\vec E\cdot\vec B)\over\partial t}=0
\ee
we obtain the (linear) constitutive relation
\be
\label{relohm}
\vec j={(\vec E\times\vec B)\nabla\cdot\vec E+
(\vec B\cdot\nabla\times\vec B-\vec E\cdot
\nabla\times\vec E)\vec B\over\mu_0 B^2}
\ee
which can be substituted into Eq.~(\ref{twomax}), (\ref{forcefree}) 
to obtain a five dimensional
(using $\nabla\cdot\vec B=0$) hyperbolic set of equations for the
electromagnetic field \cite{uch97}
which can be solved numerically
\cite{kom01}. Generalizing to a curved spacetime, specifically that 
associated with a Kerr hole, presents no difficulties of principle. Boundary
conditions have to be specified on conducting surfaces and at the horizon.
The latter is tantamount to requiring that the electromagnetic
field be non-singular when measured in an infalling frame
\cite{zna78}. 
\subsection{Relativistic MHD}
It is instructive to contrast this electromagnetic approach with the
relativistic MHD formalism that is currently used by most investigators
\cite{phi82}, \cite{cam86}, \cite{tak90}, \cite{tak00}, \cite{par02}.
In the MHD formulation, which is almost certainly 
required for field lines that connect
to the accretion disk and the plasma surrounding the outflow, the force-free
equation must be modified to include inertial terms and, 
perhaps, pressure gradients.
This means that a fluid velocity field must be tracked, along with an 
enthalpy density and a pressure. In addition, it is generally supposed that 
the current density satisfies an Ohm's law in the rest frame of 
the fluid.
\be
{\vec j\over\sigma}=\vec E+\vec v\times\vec B\rightarrow0
\ee
Under relativistic MHD,
the constitutive relation, Eq.~(\ref{relohm}) must be augmented with inertial
terms. The equations are still evolutionary, though more complex.

The introduction of these extra complications, when the inertia
of the plasma is relatively small, can be questioned 
on several grounds. Firstly, it is assumed that the electric field 
vanishes in the center of momentum frame of the plasma. This is not 
guaranteed by the plasma physics. Secondly, it is usually 
assumed that the plasma slides
without friction along the magnetic field. In other words, there is no 
dissipation. This is unlikely to be the case
in the face of instabilities and radiative drag
\cite{bes93}. Thirdly,
it is generally supposed that plasma is conserved. This is untrue of 
a black hole ergosphere, where pair creation 
is going on all the time.  Finally, there
is the assumption that the particle pressure tensor is isotropic. 
In practice this is rarely the case in observed plasmas, including the
solar wind. This is relevant to the discussion of the sound speed
and is of crucial importance to the asymptotic behaviour of characteristics. 
\subsection{Gaps}
There is a potential problem with both the force-free
and the relativistic MHD approaches. This was uncovered in the original, 
Goldreich-Julian\cite{gol69} 
model of an axisymmetric pulsar. The sign of the space charge
in a force-free magnetosphere is given by the sign of $\vec\Omega\cdot\vec
B$ and can change along a magnetic field line. This means that positive 
current must stream through outward regions of negative charge density
(or {\em vice versa}). This, in turn, implies that positive and negative 
charges have a substantial relative velocity (as required in some models
of pulsar emission). These conditions may lead to the the formation of 
``gaps'', within which enough fresh pairs are created
to allow these conditions to be satisfied. 

Actually, there is no kinematic requirement that any pairs be created
to satisfy the current and charge density change in any stationary 
outflow as long as the density is large enough. The currents 
of the positrons (or protons) and electrons 
can be separately conserved and it is always possible to adjust the 
mean velocities of these two species so as to produce any required
charge density \cite{bla76}. However it is not clear 
that the plasma will cooperate
electrodynamically when we include inertia and instabilities. If, for example,
all the particles travel ultrarelativistically with  
Lorentz factors $\gamma>>1$, then either the particle density
must exceed the Goldreich-Julian density by a factor
$=O(\gamma^2)$ or there will be pair production in a gap. Under
these circumstances there may well be 
time-dependence, dissipation and emission of radiation, as has 
been proposed in the case of pulsars. 

Despite this concern, the potential differences 
required to create pairs through a vacuum breakdown are typically $\sim1$~GV
and never more than $\sim1$~TV. These are orders of magnitude smaller
than the EMFs required to account for ultrarelativistic 
outflows ($\sim30$~PV for the Crab pulsar, $\sim100$~EV for a powerful
AGN jet and $\sim10$~ZV for a long GRB). Therefore, 
I shall assume that the plasma solves these problems without expending
much of its motional EMF along the magnetic field lines or invalidating
the force-free approach inside magnetospheres.
\subsection{Currents and Fields}
Electromagnetic models exhibit some simple generalities that derive
from treating them as electromagnetic circuits. The source region can be 
thought of as a generator capable of sustaining an EMF $\cal E$. Under most
conditions the maximum energy to  which a particle can be 
accelerated will be limited by $\sim Ze{\cal E}$ and the load will be
electromagnetic implying that its effective impedance $Z_{{\rm load}}$
is typically that of free space $Z_{{\rm load}}\sim100\Omega$.
This implies that the power lost to the load
is $\sim{\cal E}^2/Z_{{\rm load}}$. For example if an electromagnetic
source is invoked to account for the ultra high energy cosmic ray protons,
an EMF ${\cal E}\sim0.3$~ZV must be invoked and a quasar-like power
of $\sim10^{46}$~erg s$^{-1}$ will usually be present. This is not an 
invariable law, but a necessary feature of most particle acceleration 
models. (Of course as we discuss below, this power is not necessarily 
radiated, it might appear as bulk kinetic energy or heat.) 
\subsection{Characteristics}\label{subsec:charact}
In order to analyze the causal behaviour of electromagnetic models
of ultrarelativistic outflows, it is necessary to consider the 
characteristics. These are simplest to consider in the 
zero inertia, force-free case. There are then just two 
characteristics, a fast mode and an intermediate mode. 

The fast mode 
is easiest to discuss in the (primed) frame in which the background
electric field vanishes. In this frame, it is simply 
an electromagnetic wave polarised such that its electric 
vector is parallel to $\vec k'\times\vec B'$. Transforming into a general,
unprimed frame we find that the common phase and group velocity
is $\vec V_f=\hat{\vec k}$ and that the electric and 
magnetic perturbations are given by
\bea
\delta\vec B&=&A\vec V_f\times[\vec E+\vec V_f\times\vec B]\\
\delta\vec E&=&A\vec V_f\times[\vec B-\vec V_f\times\vec E]
\eea
where $A$ is a dimensionless amplitude \cite{ucb97}.
Note that the electric and magnetic perturbations are equal in magnitude
and form an othogonal triad with the wave vector $\vec k$. 
The quantity $|\delta\vec B|/\omega$ is Lorentz invariant.

The intermediate mode
is also best understood in the primed frame.
The electric and magnetic perturbations are also equal in 
magnitude and form an orthogonal triad with $\hat{\vec B'}$ 
so that the group velocity is $\hat{\vec B'}$, (assuming that 
$\vec k\cdot\vec B>0$, without loss of generality). When we 
transform to the unprimed frame, the group velocity becomes
\be
\vec V_i={\vec E\times\vec B+\left(B^2-E^2\right)^{1/2}\vec B\over B^2}.
\ee
This is what is important for the propagation of information. 
The electric and magnetic perturbation are still equal and 
form an orthogonal triad with $\vec V_i$ in this frame, so that
\bea
\delta\vec B&=&{AB\over\omega}\vec k\times\vec V_i\\
\delta\vec E&=&\delta\vec B\times\vec V_i
\eea
This time, the quantity $(1-E^2/B^2)^{1/2}|\delta\vec B|$ is Lorentz 
invariant.

Next, consider a force-free electromagnetic field rotating
with angular velocity $\Omega$ 
in flat space so that $\vec E=-(\vec\Omega\times\vec r)
\times\vec B$.  There are two intermediate group velocities
with components parallel and antiparallel to $\vec B$. A calculation 
shows that the inward-directed
group velocity has a radial component that changes from negative to positive 
at the {\em intermediate critical surface}, (also called the light cylinder),
where $|\vec\Omega\times\vec r|=1$. By contrast,
in a stationary axisymmetric, black hole magnetosphere, where spacetime 
is curved, there are two intermediate critical surfaces, 
an inner one, within which outward-directed intermediate
modes cannot propagate out, and an outer one beyond which inward-directed
modes cannot propagate in. There are also fast critical surfaces. 
The innner one coincides
with the horizon; the outer one is located at infinity.
The addition of a small amount of inertia,
under the rubric of relativistic MHD, changes this, in ways that have proven
to be controversial
\cite{oka01}. It leads to the formation of fast mode critical
surfaces at finite locations beyond which the electromagnetic field 
is isolated. However, providing the inertial loading is small, this should
not lead to a large change in the magnetospheric structure. 

Now, consider the information carried by these waves in the far field
under the force-free approximation.
The intermediate modes can only communicate information outward, beyond their
critical surface. However, the group velocities of
fast modes can point in all directions, including inward. 
If the background magnetic field is nearly toroidal
and $\vec k$ is poloidal, (as it will be for axisymmetric 
disturbances), then the magnetic
perturbation is also almost toroidal. As such it carries information about the 
poloidal current and allows inward communication 
of this information beyond the light cylinder.
(It is not necessary for the wave to include an electrical current 
perturbation for this to happen \cf \cite{pun01}.) 
Similar considerations apply in a force-free 
black hole magnetosphere within the inner, intermediate critical surface.
\section{Electromagnetic Extraction of Energy from Spinning Black Holes}
\subsection{Direct Extraction}
\begin{figure}[b!]
\begin{center}
\includegraphics[width=.8\textwidth]{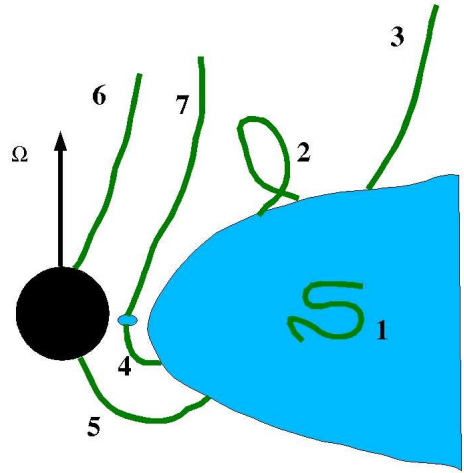}
\end{center}
\caption{Different types of magnetic field line discussed in the text. 
1. The interior torque is contributed by magnetic field amplified through 
the magnetorotational instability. 2. Short loops of toroidal field
will energise the disk corona, through having their footpoints being twisted
in opposite senses and creating small scale flares. 3. Open field lines
that connect the disk to the outflow may drive a hydromagnetic wind. 
Loops of field from the inner disk that connect to plunging gas (4.)  or the 
event horizon (5.) of the hole can
effectively remove energy from the hole. 
6. Open field lines that cross the event horizon can power a relativistic 
jet, which may be collimated by and possibly decelerated by the 
outflow. 7. Open field lines that connect to plunging gas in the
ergosphere may likewise contribute to the jet power.}
\label{fig:fields}
\end{figure}

There are several ways through
which the rotational energy associated with the spinning spacetime can be 
tapped electromagnetically \cite{bla77}, \cite{phi82}, \cite{mac84},
\cite{sue85}, \cite{tho86}, \cite{zha89},
\cite{lee00}, \cite{van01}, \cite{bla01}, Fig.~\ref{fig:fields}. The 
particular choice that I have emphasised,
because I believe that it represents the dominant energy channel, is that
the horizon is threaded by a large flux of open magnetic field, supported
by external electrical currents flowing in the inner accretion disk. The 
magnetic flux may have accumulated as a result of being carried 
inward by the inflowing gas or as result of dynamo action in the disk
\cite{kud96}.
(These two processes are unlikely to be cleanly separable.)

However, if flux does thread the hole, (in some appropriate 
infalling coordinate system \cite{kom01}), a continuous,
electromagnetic Penrose process operates in the ergosphere of the black hole
which results in Poynting flux flowing inward across the horizon 
and, simultaneously, propagating away from the hole to infinity. 
How can this be? The answer is that that energy flux is 
conserved in a frame at rest with respect
to Boyer-Lindquist coordinates, which becomes singular
at the horizon. Therefore power appears to emerge from the 
horizon in the Boyer-Lindquist frame. 
However, physical observers must orbit with respect 
to this coordinate system. Doppler boosting the energy flux into a frame 
moving with a physical observer produces an inwardly directed energy flux,
as we expect.
The source of the power is ultimately the spacetime, 
against which the electromagnetic field in the ergosphere
is doing work.

Another concern is that the current-carrying charged particles,
presumably pairs, should flow inwards at the horizon and presumably 
outwards, far from the hole. This requires pair creation in the region
between the ingoing and outgoing light (intermediate) surfaces. As discussed
above, this will happen quite readily either through cross-field diffusion,  
pair production by ambient $\gamma$-rays or by opening up a gap
\cite{his01}. The ratio of the minimum pair density required to carry the
current to the maximum that could be present without inertia being important
is of order the ratio of $\Omega_H$ to the electron gyro frequency
$\sim10^{-15}$ in the context of a powerful AGN.

The net efficiency of energy extraction can be considered using a circuit 
analogy. The Poynting flux that flows into the black hole (in a frame
rotating with angular velocity $\Omega_H$ can be considered as a 
form of dissipation or internal resistance. Likewise, the Poynting
flux far from the hole represents a second (load) resistance.  The efficiency
of energy extraction -- the fraction of the reducible mass that is actually
dissipated in the load -- depends upon the angular velocity of the 
field lines. In a stationary, axisymmetric electromagnetic model, 
this is a resistance-weighted average of the hole and load angular velocities.
It depends upon the shape of the magnetic surfaces far from the hole.
For example, for a slowly rotating hole,
with a radial field, the efficiency is 0.5; when 
the magnetic surfaces are paraboloidal, the average efficiency is 0.38.
Collimating outflows are generally less efficient.

However, not all the field lines that thread the horizon need connect with the
outflow. Some low latitude field lines may connect directly to the accretion 
disk and provide a supplementary power source for the disk as well as 
a possible driving torque for exciting quasi-periodic oscillations.
\cite{bla99}, \cite{lip00}. 
This energy channel could be important,
especially if the disk is thick. However, it is unlikely to lead to 
an ultrarelativistic outflow. 

The process that I have just described -- the direct electromagnetic
extraction of energy from the black hole -- is distinct from 
(though can operate
simultaneously with) an alternative process, the extraction of binding energy 
from the accreting gas both in the disk \cite{bla76}, \cite{bla82}, 
\cite{lee01} and in the plunging region
between the inner edge of the disk and the horizon \cite{koi00}, \cite{mek01}.
The extra power 
that this process produces can be charged to the spin energy of the hole,
which increases at a slower rate than it would do so in the absence 
of magnetic stress. However
the intermediate working substance that effects this transformation
is the inertia of the plasma not the electromagnetic field. 
One way of accounting for the greater energy release of power from
an accreting, Kerr hole is that it is ``borrowed'' 
from the extractable spin energy of the hole. After the 
gas eventually crosses the horizon the hole mass increases by a smaller
amount than it would have done in the absence of the magnetic stress.
(The magnetic connection of the disk to the 
plunging gas seems to be a less promising source of 
power because the magnetic field lines which are likely to have a far
greater inertial load than open field lines threading the hole,
either quickly reconnect or become super-fast, leaving the inflowing  
gas effectively disconnected from the disk \cite{hir92}, \cite{ago00},
\cite{gam99}, \cite{arm01}.)

There are three 
reasons for emphasising direct extraction of energy from the
hole over extraction from the infalling gas at least for a rapidly 
spinning hole. The first is that the event horizon has a larger effective
area than the annular ring between the hole and the disk. The second is that
any gas-driven outflow is likely to be contaminated with 
baryons and consequently
is unlikely to achieve the ultrarelativistic outflow velocity
required.  The third is that holes probably rotate
much faster than orbiting gas, except quite close to the horizon, from where
the extraction of energy will be quite inefficient.
\subsection{Criticisms and Variations}
The general idea that the spin energy of a black hole can be tapped
electromagnetically as just
discussed, has received an observational boost from the discovery 
that black holes are commonplace on both the stellar and the massive scale
(as well as, perhaps, on the intermediate scale
\cite{col99}) and that the second parameter,
the spin, appears to be 
large so as to allow gas to orbit close to the horizon and to
form strongly redshifted emission lines \cite{tan95},
\cite{wil01}. However, the more specific 
notion that ultrarelativistic outflows are powered
by direct, electromagnetic extraction of energy from a spinning hole
has been criticized on several grounds and some alternative
models have been developed.  These criticisms and alternatives include,
in addition to those discussed above: 
\begin{itemize}
\item Black hole magnetospheres develop 
gravitationally significant space charge particularly just after the 
formation of the event horizon. The foregoing considerations
imply that it is necessary for large, field-parallel electric fields to 
develop\cite{ruf01}, \cite{van01}, \cite{pun01}.
In the limiting case the potential differences along the magnetic field lines
approach the impressively large, fundamental ``gravitational'' value, 
$(G\mu_0)^{1/2}c^3\sim1$~XV. (1 xenna eV $\equiv10^3$ yotta eV 
$\equiv10^{27}$~eV!)
\item The electromagnetic field has a vacuum configuration with
$\vec E\cdot\vec B\ne0$. The potentials here are significantly smaller
$\sim GMB/c$ though still comparable with the full EMF developed in a 
force-free model, as discussed above and, on the face of it, ample to ensure
discharge of the vacuum. If an electromagnetic
configuration like this can be sustained then interesting, new physical 
effects can be contemplated \cite{tom01}, \cite{hey01}.
\item The stationary, force-free, electromagnetic configuration 
is unstable because the region within the
inner light surface is effectively out of causal contact with the horizon
and the magnetosphere cannot respond to changes in the spin of the hole
\cite{pun01}.
Probably the most convincing way to decide if this really
happens is to perform 
time-dependent numerical calculations of a black hole magnetosphere
and determine if it settles down to a stationary state with 
electromagnetic energy being steadily extracted. Preliminary 
computations show no sign of instability \cite{kom01}, \cite{tom01}.
\item A rather different type of
instability is to non-axisymmetric screw modes. These are likely
to be present especially when the magnetic field becomes tightly 
wound as we discuss further below. What is important is what happens
in the nonlinear regime.
\item The force-free equations are fundamentally inadequate
and only relativistic MHD can provide a complete description 
of a magnetosphere and its outflow particularly when 
considering the role of critical points \cite{phi82}, \cite{oka01}.
\item The angular velocity of the magnetic field lines
is determined by the details of the pair creation not the boundary
conditions at the horizon and large radius. If so, the efficiency
of energy extraction 
is expected to be seriously reduced \cite{bes01}, \cite{phi82}, \cite{pun01}.
\item Electromagnetic extraction of energy from the hole is 
irrelevant because too little flux threads
the horizon for the power extracted to be significant relative to the power
extracted from the disk \cite{ogi99}.  This is undoubtedly the case for slowly 
rotating holes with thin disks that only extend down to $r\sim6m$
\cite{bla77}. It will not be clear what is the disk - hole power ratio
for rapidly spinning holes
until we have a better understanding of disk structure and solve
the cross-field stress equation in the magnetosphere. Note that it is still
possible for the jet power to exceed the bolometric luminosity 
and yet be less than the total power lost by the accreting gas if the 
disk does not radiate efficiently, which is commonly assumed to be the
case for radio galaxies and quasars \cite{mei01}.
\end{itemize}

Many of these issues are unlikely to be satisfactorily resolved 
until there are an extensive series of time-dependent numerical 
simulations using either the force-free or the relativistic 
MHD formalism, or preferably both \cite{koi00}, \cite{kom01}, \cite{mei01}.
\section{Electromagnetic Shells}
\subsection{Electromagnetic Solution}
\begin{figure}[b!]
\begin{center}
\includegraphics[width=.8\textwidth]{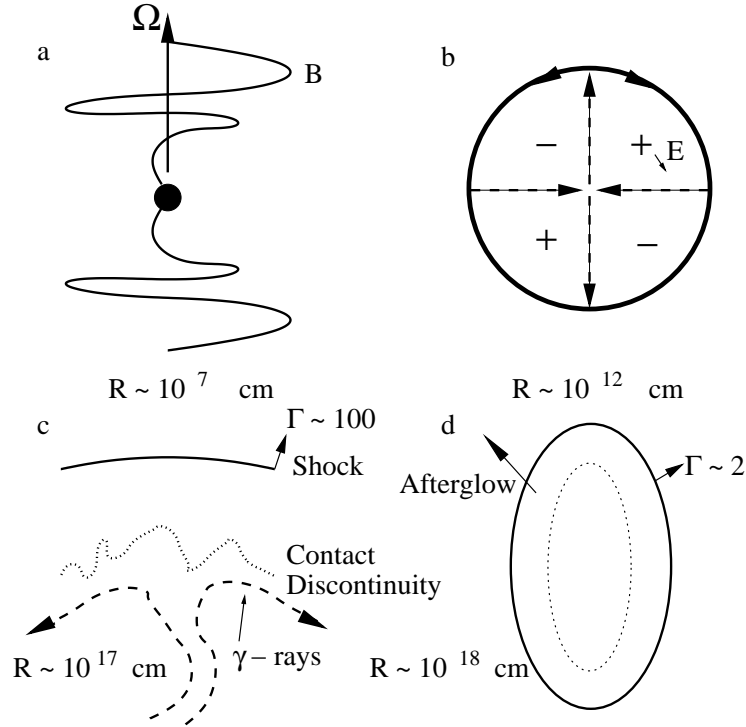}
\end{center}
\caption{Four stages in the expansion of a magnetic shell with scales
appropriate to a long duration GRB.
a) The magnetic field $\vec B$ changes from poloidal to
toroidal close to the outgoing light surface of the magnetic rotator
at a radius $R\sim10^6$~cm. The alternating component of the electromagnetic
field decays relative to the DC toroidal field. b) The source is active 
for $\sim100$~s. By this time, it will have inflated a magnetic
bubble with radius $R\sim3\times10^{12}$~cm, expanding with Lorentz factor
$\Gamma\sim3\times10^4$.The magnetic field is mostly toroidal, with the
signs shown, while the electric field $\vec E$ is poloidal. The quadrupolar 
current flow is shown dashed. The shocked circumstellar medium
is compressed into a thin shell of thickness $\sim10^3$~cm.
c) By the time the shell has expanded to $R\sim10^{17}$~cm,
$\Gamma\sim100$ and most of the 
electromagnetic pulse has caught up with the blast wave. This phase
is observed a time $\sim100$~s after the initial explosion. The current flow
is still largely quadrupolar, though it is unstable along the axis 
and the equator and this drives an electromagnetic turbulent cascade.
which ultimately creates electrical resistance and dissipation in the form
of pair production, particle acceleration and intermittent, 
$\gamma$-ray emission. These instablities also promote corrugation of the
contact discontinuity and incorporation of the magnetic field into
the shocked interstellar medium where it can mix with relativistic electrons
accelerated at the bounding shock front. d) When the blast wave has expanded
by a further factor $\sim30$, its speed is only mildly relativistic. Its shape
will be quite prolate as the expansion is fastest along the axis. Most of 
the energy released by the central, spinning, magnetic rotator is now carried
by the shocked interstellar medium.}
\label{fig:magshell}
\end{figure}

Suppose that a magnetic rotator - black hole plus torus or a 
magnetised neutron star spins off
magnetic flux for a time $t_{{\rm source}}$ into the 
far field,
and that this creates a relativistically expanding shell 
of electromagnetic field 
that drives a blast wave into the surrounding medium. The blast wave is
supposed to be bounded on its outside by a strong shock front that moves with 
Lorentz factor $\Gamma$ and, on its inside, by a contact discontinuity,
separating it from the electromagnetic shell, 
that moves with Lorentz factor $\Gamma_c$.   
Suppose further, for simplicity, that the conditions near the 
light cylinder are such that the current 
well beyond the light cylinder flows along the axes, then flows
along the contact discontinuity and finally returns 
to it source through the equatorial plane (Fig.~\ref{fig:magshell}),
just as we dicsussed above in the context of the Crab Nebula and the 
solar wind. 

If we ignore the poloidal component of the magnetic field, (and
consequently the flux of angular momentum),
then the relevant solution of the force-free equations associated 
with this current flow has the form 
\bea
\label{shell}
B_\phi&=&{f_+(t-r)+f_-(t+r)\over r\sin\theta}\\
E_\theta&=&{f_+(t-r)-f_-(t+r)\over r\sin\theta}
\eea  
The two terms in each expression are fast modes propagating outward and inward.
The charge and current density vanish in the interior
of the shell. We can determine the functions 
$f_+,f_-$ by specifying the electromagnetic field
at some small radius beyond the light surface and by matching to an 
ultrarelativistic blast wave expanding into the surrounding medium. 
This last requires that the outer surface of the shell move at the same 
speed as the inner surface of the blast wave and that the magnetic stress 
normal to this surface match the pressure in the blast wave.

The simplest assumption to make is that the
strength of the magnetic rotator is constant, ($f_+={\rm const}$),
for $t\lo t_{{\rm source}}$
and that the external
density is constant in radius. These assumptions imply that the 
Lorentz factor of the blast wave's outer shock front varies with radius 
$R$ according to $\Gamma\propto\csc\theta R^{-1/2}$
\cite{blm76}. 
(The
Lorentz factor of the contact discontinuity, $\Gamma_c$, exhibits a similar
variation.) The 
electromagnetic velocity (the velocity of the frame in which the 
electric field vanishes) in the body of the shell
$\vec E\times\vec B/B^2$ is radial and is equal in magnitude
to $(f_+-f_-)/(f_++f_-)$ and the magnetic stress in a frame
moving with this velocity is $\propto f_+f_-\csc^2\theta$.
The expansion of the blast wave 
is anisotropic, being faster along the poles,
giving an electromagnetic power per steradian $L_\Omega\propto\csc^2\theta$.

It is instructive to  relate this simple solution to our earlier 
discussion of characteristics. The backward-propagating, reflected wave
is a fast mode but
has a small amplitude when $\Gamma_c>>1$.  (In this approximation,
the intermediate mode's group velocity is purely toroidal and carries
no radial information.) However the fast mode 
is able to propagate radially inward and, in effect, create
an electromagnetic pressure wave which decelerates the electromagnetic 
velocity of the outflow and reduces it to match that of the contact 
discontinuity. The magnetic field is toroidal and carries information 
about the current flow. Put another way, if we were to change the 
properties of the load, \eg by encountering a sudden increase in the ambient
gas density, which would cause sudden jumps in $\Gamma,\Gamma_c$,
then the boundary conditions on the interior flow would change, along with 
changes in the 
amplitude of the reflected wave. This allows the interior solution to adjust. 

However, it takes a
very long time for a wave to be reflected by the blast wave and return
to the origin. This is generally true of ultrarelativistic flows
and stationary solutions, which take a long time to be established,
can be quite misleading. This has a second consequence.
There is no necessity to destroy magnetic flux through ohmic
dissipation, until the wave can actually propagate back to the source.
(This is in contrast to what happens
with PWN, where it is necessary to destroy most of the 
magnetic flux.)
Stated another way, there {\em need} be little resistance in 
the electrical circuit. The effective load can 
consist of the performance of work on the expanding blast wave. This is 
where most of the power that is generated by the central magnetic
rotator ends up. Until the blast wave becomes nonrelativistic, 
the distinction between the
inertial load of this solution and a different,  
dissipative load is quite unimportant for the behavior of the 
magnetic rotator. As long as the expansion 
remains ultrarelativistic, it is a very good approximation 
to impose a Sommerfield radiation condition on the solution of the
magnetosphere and to ignore the reflected wave. Analogous
remarks apply to the horizon boundary condition, in the specific 
case of a black hole magnetosphere..  

This simple, electromagnetic solution will only remain valid until the end of
the outward-propagating, electromagnetic pulse catches up with the 
contact discontinuity.
This occurs at some radius $R_{{\rm free}}\sim\Gamma(R_{{\rm free}})^2
ct_{{\rm source}}$. Thereafter, the surrounding blast wave which, by now,
has acquired almost all of the energy in the explosion, will expand freely 
with $\Gamma\propto R^{-3/2}$ until it becomes non-relativistic, after 
which point, the blast wave will follow a Sedov solution.
\subsection{Gamma Ray Bursts}\label{subsec:grb} 
As well as bring out some formal points, the electromagnetic solution
just described provides a possible model for GRB
\cite{mes01}, \cite{uso94}, \cite{lyu02}.
Suppose that a magnetic rotator spins off toroidal magnetic 
field as it slows down and that this magnetic field fills an 
anisotropic, relativistically expanding shell in a uniform medium.
Suppose, further, that the flux 
distribution near the light cylinder is consistent 
with the current being concentrated along the axis and in the equatorial
plane, as described above. The current density is most intense 
on the axis and, although there is no requirement that flux be destroyed as 
long as the expansion is relativistic, in practice the magnetic pinch will
become hydromagnetically unstable 
to sausage and kink modes (in the comoving frame) after expansion 
beyond a radius where the stabilising, poloidal field becomes insignificant
\cite{beg92}.
These global instabilities, which should have a longitudinal wavelength 
comparable in size to the width of the current distribution, 
may sustain an electromagnetic turbulence spectrum which should 
ultimately be responsible for particle  
acceleration and the excitation of transverse gyrational motion. 
This turbulence 
may have already been seen in the measured fluctuation power spectrum
\cite{bel00}.
One reason why particles can be accelerated is that, when the power cascades
down to short enough wavelengths, there may be too few charged particles 
to carry the electrical current and field-parallel electric fields will 
develop \cite{tho98}. In addition,
nonlinear wave-wave interactions will lead to an efficient,
stochastic acceleration \cite{bla73}. This
particle acceleration is the microscopic source of the electrical
resistance that is invoked in the global electrical circuit.

Of course the accelerated particles will also radiate in 
this electromagnetic maelstrom. There will be a mean magnetic field in 
the comoving frame which will be responsible for regular synchrotron radiation
if transverse gyrational motion is excited. However, in the absence of a
strong, high frequency, resonant turbulence spectrum, the two particle 
acceleration mechanisms identified above are more likely to pump energy
into the longitudinal motion. In this case, the emitted spectrum
is more likely to be determined by the synchro-Compton process
\cite{bla72} where the amplitudes of lower frequency wave modes 
are so strong that they cause the radiating electrons to oscillate
with angular amplitudes in excess of $\sim\gamma^{-1}$. In addition,
inverse Compton scattering is likely to produce a strong $\gamma$-ray 
spectrum above the threshold for pair production. 
All of this will take place in a frame moving
outward with the electromagnetic velocity and so the photon energies 
will suffer one final Doppler boost and the emission will be 
strongly beamed outward in the frame of the explosion. (By contrast,
no such boost will be necessary in the case of PWN jets which explains
why they can be seen from large inclination angles.)
 
Global pinch instabilities also provide a 
plausible explanation for the large $\sim1-10$~ms fluctuations in the observed 
$\gamma$-ray flux that are observed. The reason why it is possible to 
circumvent the usual argument that the emission profile will be smooth
\cite{sar98}, is that the emitting elements (essentially 
fast or intermediate wavepackets) move with relativistic speed 
in the electromagnetic frame. A distant 
observer may therefore only see the few small intense beams  at a given time
that are beamed towards him. The number and emissivity of these patches
can vary through the total burst while the radiative efficiency can remain
high. 

Actually, although the current density is strongest along the pole,
it should also be quite strong along the contact discontinuity 
at the outer boundary of the pulsar wind nebula and in
the equatorial plane. $\gamma$-ray emission could also arise as a result of 
local instabiliites from these regions as
well and this will change the predicted beaming properties. 
\subsection{Afterglows} 
The afterglow is formed after the blast wave becomes free of
its electromagnetic driver. Now, in most afterglow models, 
including those involving jets, it is supposed that the 
expansion velocity does not vary with angle. However,
an electromagnetically-driven 
blast wave necessarily creates an anisotropic explosion and this has important
consequences for observations of the afterglow, especially in the 
ultrarelativistic phase of expansion. If we continue to use our simple model,
we find that the afterglow expansion varies most rapidly and remains
relativistic for longest closest to the symmetry axis. 
As $L_\Omega\propto\csc^2\theta$ the energy contained in each 
octave of $\theta$ is roughly constant.
This means that the most intense bursts and afterglows 
in a flux-limited sample will be seen pole-on 
and can exhibit achromatic breaks when $\Gamma\sim\theta^{-1}$, 
which might be mistaken for jets.
The inferred explosion energy with our simple model
will be roughly independent
of $\theta$ and characteristic of the total energy
\cite{fra01}.
When the expansion becomes non-relativistic, the remnant will have a 
prolate shape which might be measurable.
(It is tempting to associate some of the barrel-shaped supernova remnants 
observed in our Galaxy with the remnants of electromagnetic explosions
\cite{kes87}.) 

This electromagnetic model provides a solution to the puzzle
of how to launch a blast wave that extends over an angular scale
$>>\Gamma^{-1}$ and where the individual parts are out of causal contact.
(Something similar happens in the early universe. Indeed the 
computation of the temporal fluctuations in GRB has some features in common 
with the computation of angular fluctuations in the microwave background.)  
In the electromagnetic model, the energy is transferred to the blast wave
by a magnetic shell that pushes (unevenly) on the surrounding gas
all the way out to $R_{{\rm free}}$. It also supplies an origin
for the magnetic flux in the blast wave, for which the alternative
origin in the bounding shock front seems very hard to explain
\cite{blm77}, \cite{gru01}. 
In the present model, magnetic field can simply 
be mixed into the blast wave (and the shock-accelerated
relativistic electrons) at the contact discontinuity
through instabilities, much like what seems to happen in regular supernova
remnants.
\subsection{Some Numbers}\label{subsec:numbers}
Let us give some illustrative, orders of magnitude for one model of a long 
duration GRB. The electromagnetic energy flux near the pole is 
$L_\Omega\sim10^{50}$~erg s$^{-1}$ sterad$^{-1}$ and lasts for 
a time $t_{{\rm source}}\sim100$~s. The associated EMF in the electrical
circuit $\sim10~Z$~V and the current is $\sim100~$~EA.
(A potential difference this large, made available along the 
contact discontinuity, provides one of the few astrophysical options for 
accounting for UHE cosmic rays \cite{wax95}.)
If the external density is uniform with  
$n\sim1$~cm$^{-3}$, then the blast wave is driven by the electromagnetic pulse 
with Lorentz factor $\Gamma\propto R^{-1/2}$ until $R\sim R_{{\rm free}}
\sim10^{17}$~cm, $\Gamma\sim\Gamma_{{\rm free}}\sim100$. 
Thereafter there is a freely expanding
blast wave with $\Gamma\propto R^{-3/2}$ until the expansion becomes 
non-relativistic when $R\sim R_{{\rm NR}}\sim3\times10^{18}$~cm.

If most of the GRB emission  (around $\sim1$~MeV)
is produced when $R\lo R_{{\rm free}}$ by the  synchrotron emission
then this requires $\sim100$~GeV electrons 
and a comoving magnetic field of strength 
$B\go30$~G. The comoving cooling time of these electrons is $\sim3$~s,
a fraction $\lo10^{-4}$ of the expansion timescale and so
if the $\sim100$~GeV pair energy density is maintained at a 
significant fraction of the equipartition energy density, then the magnetic
energy can be efficiently transformed into $\gamma$-rays. 
The opacity to pair production for a $\gamma$-ray of energy $E$
is $\sim0.1(E/1{{\rm GeV}})$. The Thomson optical depth depends upon 
the details of the particle acceleration but is 
plausibly much smaller than unity so that the observed $\gamma$-rays can
freely escape without erasing the variability.  
\subsection{Some Possible Generalizations}
This simple model of GRB was predicated upon a very simple current
flow. In practice, it is the detailed electrodynamics in the 
vicinity of the outgoing light surface that fixes the poloidal 
magnetic field and electrical current distributions. We can therefore 
change these (still, of course, maintaining the force-free condition)
and solve for a new evolution of the magnetic shell and blast wave. A broader 
distribution of currents will generally produce a less 
pronounced expansion along the axis and change somewhat the statistics of 
the observed afterglows. In the solution above, we ignored the poloidal 
magnetic field and, consequently, the angular momentum. These can 
be reinstated perturbatively into the solution. Their influence wanes
with increasing radius. 

Other ways to obtain different solutions
include changing the temporal variation of the source from the simple
step function considered above and allowing the external density
to vary with radius and latitude. As the individual parts of the 
blast wave expand essentially independently, when ultrarelativistic,
there are no new issues of principle to address in solving these problems.
However, all this changes when the blast wave becomes non-relativistic.
At this point the interior gas will be roughly isobaric and the shell 
will become more spherical with time.  
\section{Discussion}
In this brief overview, I have explored the strong
version of the ``electromagnetic hypothesis'' for ultrarelativistic
outflows, namely that they are essentially electromagnetic phenomena 
which are driven by energy released by spinning black holes or neutron 
stars and that this electromagnetic behavior continues into the source
region even when the flows become non-relativistic.
(If this hypothesis is falsified in
the resolved, emission regions, then there is the fallback position- that the 
outflows are only initially electromagnetic and quickly convert to a 
baryonic  -- jet \cite{vla01}.) The most striking implications of the electromagnetic hypothesis 
are that particle acceleration in the sources is due to 
electromagnetic turbulence
rather than shocks and that the outflows are cold,
electromagnetically dominated flows, with very few baryons 
at least until they become strongly dissipative.

There are many possible tests of this hypothesis. 
GRB, should not be associated with strong, high energy neutrino sources.
(A gravitational wave signal is expected in some, though not all, source
models and would be strongly diagnostic if ever detected.) It is possible, 
though not very likely, that electromagnetic
GRB will be associated with core-collapse supernovae.
In this case there may be a detectable MeV neutrino pulse.
More immediate, though less specific, diagnostics include relating the 
duration and character of the GRB with the inferred 
observation angle of the burst -- higher inclination should be associated
with longer and less intense bursts. 

The spectrum and polarisation of the 
afterglow emission might also contain some clues, though the lack of 
a usable theory of particle acceleration at relativistic shocks 
is a handicap. 
(Recent, promising progress on this problem produces
a power law distribution function with a logarithmic slope of 2.2
\cite{ach01},
provided that the scattering is essentially normal to the shock front.
What is not yet clear is whether these scattering conditions are present
or whether rather different principles \cite{gal94} might be at work.)
On general grounds, it would be of strong 
interest to determine if the particle acceleration 
properties of ultrarelativistic shocks propagating into
interstellar gas are just a function of the shock Lorentz factor 
$\Gamma$, while scaling with the external density, 
as should be the case. It is also important   
to try to understand kinematically if the magnetic field 
is introduced into the blast wave at the shock, as conventionally 
assume, or at the contact discontinuity as proposed here.
A more detailed discussion 
of the GRB emission, than presented
here should also account for the MeV breaks observed in $\gamma$-ray spectra.

From a more theoretical 
perspective, there is much to be learned about the properties of 
force-free electromagnetic fields, especially their stability. The 
possible relationship of the GRB fluctuation power spectrum to an 
underlying turbulence spectrum is especially tantalising. Undoubtedly,
numerical simulations will be crucial as the problem is essentially
three-dimensional. Force-free electromagnetism is easier to study
than relativistic MHD and may well be a very good approximation
in many of these sources.

Turning to PWN, the most direct, observational challenge is to see
if there really is a strong, dissipating shock as expected with a fluid wind
or a flow of electromagnetic energy towards the axis and the equatorial
plane as predicted by the electromagnetic model and as appears to 
be exhibited by the Crab Nebula. It is worth trying
to model the images from radio to X-ray wavelengths phenomenologically 
adopting the electromagnetic  hypothesis.  
For the pulsar, the oblique rotator remains an 
unsolved, theoretical problem and which can now be tackled numerically,
at least in the time-dependent, electromagnetic limit
\cite{con99}. If a satisfactory,
global, electromagnetic solution can be found, then this provides 
a framework for carrying out microphysical investigations to revisit such 
question as to  how currents manage to change their space charge density 
as they flow along magnetic field lines. 

Finally, for AGN, it is predicted that the jets are not hadronic
and should also not be powerful high energy neutrino
sources. This is refutable. In addition, we need to see if the well-resolved 
jets and their lobes can be re-interpreted in terms of an 
unstable Z-pinch. A good 
place to start is through mapping the 
magnetic field using polarisation measurements and the
internal mass density through internal depolarisation data.
In addition, detailed imaging spectra
from radio to X-ray energies can be used to determine where the 
particles are being accelerated -- at shock surfaces or in volumes
containing strong, unstable currents -- and how they diffuse away from
these acceleration sites at different energies.

The prospects are pretty good for determining if cosmic lighthouses 
are powered by electricity or fire. 
\section*{Acknowledgments}
I thank the organisers of this workshop for their hospitality
and patience and Mitch Begelman, Paolo Coppi,
Arieh K\"onigl, Amir Levinson,
Max Lyutikov, Chris McKee, David Payne, Martin Rees and Roman Znajek for their 
collaboration. Brian Punsly and 
Maurice van Putten are also thanked for stimulating debates
on some of the issues raised here. 
Support under NASA grant 5-2837 is gratefully acknowledged.

\end{document}